\documentstyle[epsfig]{mn}
\begin{document}
\LARGE
\normalsize

\title[MERLIN observations of GRS 1915+105]
{MERLIN observations of relativistic ejections from \\ GRS 1915+105}
\author[R.P. Fender et al. ] 
{R.P. Fender$^{1}$\thanks{email : rpf@astro.uva.nl},
S.T. Garrington$^{2}$,  D.J. McKay$^{2,3}$,
T.W.B. Muxlow$^{2}$, \cr G.G. Pooley$^{4}$,
R.E. Spencer$^{2}$, A.M. Stirling$^{2}$, E.B. Waltman$^{5}$
\\
$^{1}$ Astronomical Institute `Anton Pannekoek' and Center for High Energy
Astrophysics, University of Amsterdam, Kruislaan 403, \\
1098 SJ Amsterdam, The Netherlands. \\
$^{2}$ University of Manchester, Nuffield Radio Astronomy Laboratories,
Jodrell Bank, Cheshire, SK11 9DL \\
$^{3}$ Joint Institute for VLBI in Europe, Postbus 2, 7990 AA Dwingeloo,
The Netherlands\\
$^{4}$ Mullard Radio Astronomy Observatory, Cavendish Laboratory,
Madingley Road, Cambridge CB3 0HE\\
$^{5}$ Remote Sensing Division, Code 7210, Naval Research Laboratory,
Washington, D.C. 20375-5351, USA\\}

\maketitle

\begin{abstract}

We present high resolution MERLIN radio images of multiple
relativistic ejections from GRS 1915+105 in 1997 October /
November. The observations were made at a time of complex radio
behaviour, corresponding to multiple optically-thin outbursts and
several days of rapid radio flux oscillations. This activity followed
$\sim20$ days of a plateau state of inverted-spectrum radio emission
and hard, quasi-stable X-ray emission. The radio imaging resolved four
major ejection events from the system.  As previously reported from
earlier VLA observations of the source, we observe apparent
superluminal motions resulting from intrinsically relativistic motions
of the ejecta. However, our measured proper motions are significantly
greater than those observed on larger angular scales with the
VLA. Under the assumption of an intrinsically symmetric ejection, we
can place an upper limit on the distance to GRS 1915+105 of $11.2 \pm
0.8$ kpc. Solutions for the velocities unambiguously require a higher
intrinsic speed by about 0.1$c$ than that derived from the earlier VLA
observations, whilst the angle to the line-of-sight is not found to be
significantly different.  At a distance of 11 kpc, we obtain solutions
of $v = 0.98_{-0.05}^{+0.02}c$ and $\theta = 66 \pm 2$ degrees. The
jet also appears to be curved on a scale which corresponds to a period
of around 7 days.

We observe significant evolution of the linear polarisation of the
approaching component, with large rotations in position angle and a
general decrease in fractional polarisation. This may be due to
increasing randomisation of the magnetic field within the ejected
component. We do not at any time detect significant linear
polarisation from the core, including periods when the flux density
from this region is dominated by radio oscillations. The power input
into the formation of the jet is very large, $\geq 10^{38}$ erg
s$^{-1}$ at 11 kpc for a pair plasma. If the plasma contains a cold
proton for each electron, then the mass outflow rate, $\geq 10^{18}$ g
s$^{-1}$ is comparable to inflow rates previously derived from X-ray
spectral fits.

\end{abstract}



\section{Introduction}

\begin{figure*}
\leavevmode\epsfig{file=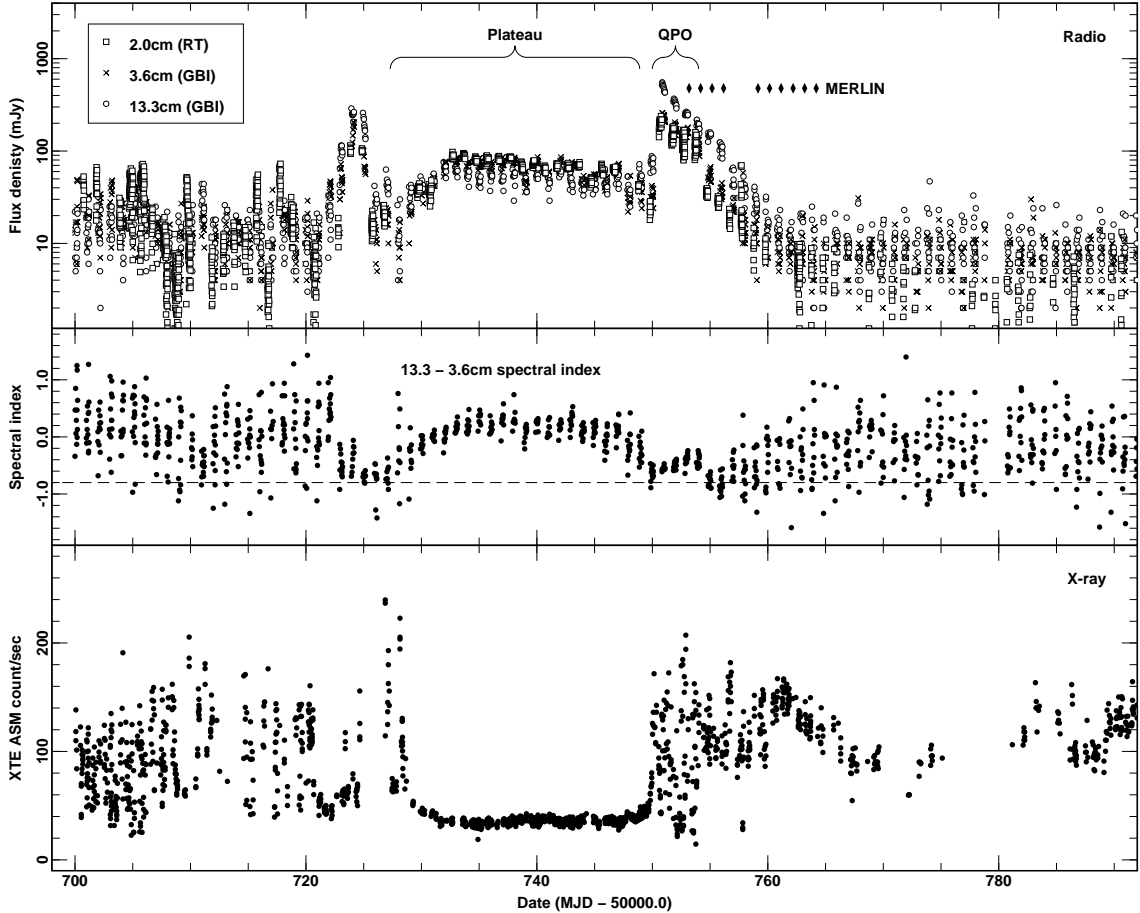,width=16cm,angle=0,clip}
\caption{Radio and X-ray flux monitoring spanning the sequence of
radio ejections mapped by MERLIN.  The upper panel shows the flux
densities measured at 2.0 cm with the RT, and at 3.6 cm and 13.3 cm
with the GBI.  The middle panel shows the radio spectral index from
simultaneous measurements at 13.3 and 3.6 cm. The lower panel shows
the count rate in the 2-12 keV band by the Rossi X-ray Timing Explorer
(RXTE) All-Sky Monitor (ASM). The optically thin state evolves towards
a spectral index of $-$0.8 which is shown by the dashed line in the
middle panel.}
\end{figure*}

\begin{figure}

\leavevmode\epsfig{file=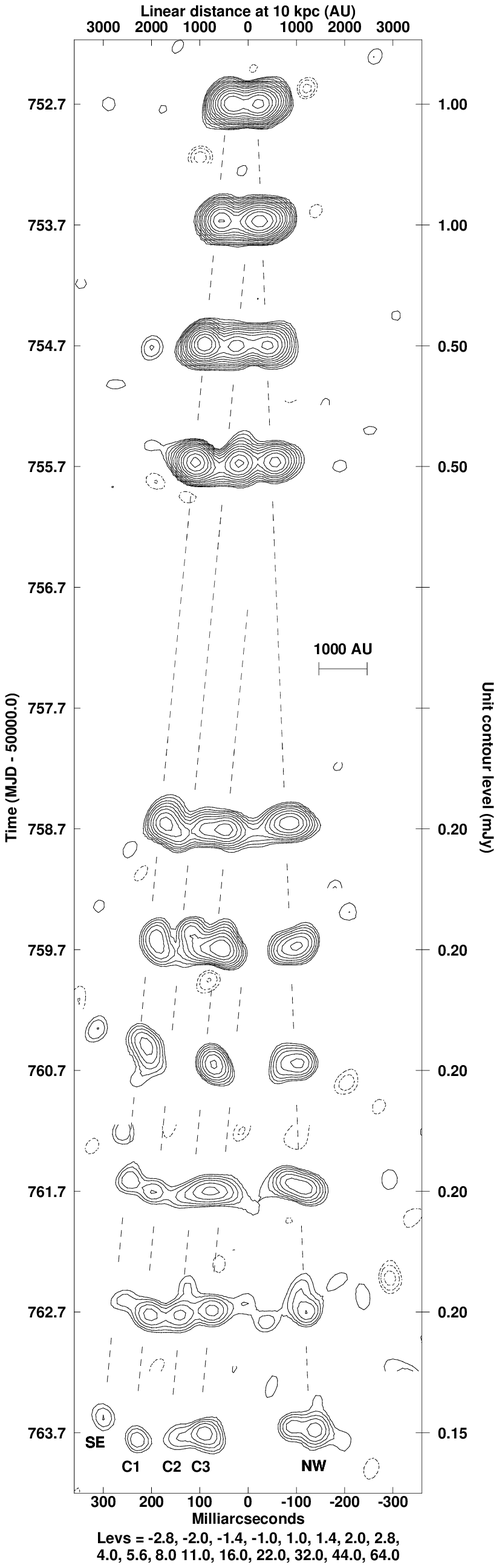,width=6.95cm,angle=0,clip}
\caption[]{MERLIN maps of total intensity from 10 -- 12--hr observations
at the epochs illustrated on Fig. 1.  The images have been rotated
clockwise by 52 degrees to form the montage. Contour levels increase
in factors of $\surd$2 from the unit contour level indicated at the
right--hand side of each image. Mapping fidelity is dominated by
dynamic range considerations for the first 5 epochs, whilst receiver
noise dominates later.}

\end{figure}

GRS 1915+105 is a distant black-hole-candidate X-ray transient in the
Galactic-plane, discovered by the WATCH instrument on board the GRANAT
mission in 1992 (Castro-Tirado, Brandt \& Lund 1992). Owing to large
extinction ($A_{\rm V} \geq 20$ mag), the properties of the optical
counterpart are unknown, but observations at radio, near-infrared and
X-ray energies have shown GRS 1915+105 to be a highly unusual and
energetic system.

Shortly after the discovery of the X-ray source, variable radio and
infrared counterparts were identified (Mirabel et
al. 1994). Subsequent mapping of the radio counterpart following an
outburst in 1994 revealed apparent superluminal motions in a two-sided
ejection of synchrotron-emitting components from the source (Mirabel
\& Rodr\'\i guez 1994; hereafter MR94). This was the first observation
of superluminal motions in our Galaxy, the only previously measured
proper motions of ejections from X-ray binaries (SS 433 \& Cyg X-3)
implying velocities of $\sim0.3$$c$. Within a year, a second
superluminal source, GRO J1655$-$40, had been discovered (Tingay et
al. 1995; Hjellming \& Rupen 1995). A third possible superluminal jet
source, XTE J1748$-$288, has recently been reported (Rupen, Hjellming \&
Mioduszewski 1998).

The VLA observations of GRS 1915+105 by MR94 revealed proper motions
of $17.6 \pm 0.4$ and $9.0 \pm 0.1$ mas d$^{-1}$ for two components
separating from a core at position angles of 150 and 330$^{\circ}$
respectively. From HI absorption measurements these authors derived a
most likely distance to the source of $12.5 \pm 1.5$ kpc, giving
apparent transverse motions of $1.25c \pm 0.15c$ and $0.65c \pm 0.08c$
respectively. Proper motion and flux density ratios between the
(inferred) approaching and receding components were consistent with
simple ballistic bulk motions. Solving for the proper motions at a
distance of 12.5 kpc, a true velocity of $0.92c \pm 0.08c$ at an angle
of $70 \pm 2^{\circ}$ to the line-of-sight was derived. Further
observations with the VLA following a radio outburst in 1995 revealed
relativistic ejection for a second time (Mirabel et al. 1996). However
the proper motions and flux density ratios appeared to be somewhat
different from those reported in MR94 possibly implying a change in
the jet velocity or the angle to the line-of-sight. A summary of
multiple VLA observations of major relativistic ejections from GRS
1915+105 is given in Rodr\'\i guez \& Mirabel (1999; hereafter RM99).


Pooley \& Fender (1997; hereafter PF97) report $\sim 2$ yr of
radio monitoring of GRS 1915+105 at 15 GHz with the Ryle Telescope
(RT).  They describe in detail quasi-periodic radio oscillations with
periods typically in the range 20 -- 40 min, a phenomenon not
previously seen in any other source, and first reported in Pooley
(1995). Fender et al. (1997) report the discovery of infrared
counterparts to the radio oscillations, suggesting an approximately
flat synchrotron spectrum from cm to $\mu$m wavelengths. They proposed
that each oscillation corresponds to a small ejection of material from
the GRS 1915+105 system. Around the same time Belloni et al. (1997a,b)
proposed that X-ray dips with similar periods correspond to the
removal of the inner ($\leq 200$ km) region of the accretion disc,
possibly advected into the black hole. The observation of a
correlation between such X-ray dips and the rise of radio oscillations
(PF97) seemed to complete a picture of repeated ejection and refill of
the inner accretion disc, the ejecta emitting a flat synchrotron
spectrum from 2 cm -- 2 $\mu$m before they cool due to adiabatic
expansion losses. Further X-ray, radio and infrared timing
observations (Eikenberry et al. 1998; Mirabel et al. 1998; Fender \&
Pooley 1998) appear to confirm this model, at least qualitatively.

Long-term comparison of radio and 2-12 keV X-ray light curves in PF97
revealed an unusual period of activity in 1996 July-August, hereafter
the `plateau' state, during which time the X-ray emission entered a
hard, quasi-stable state, while the radio emission became relatively
bright at 15 GHz. Earlier occurences of this state are reported in
Foster et al. (1996) and Harmon et al. (1997).  Bandyopadhyay et
al. (1998) also discuss bright infrared emission during such periods.
It was speculated in PF97 (and earlier, in Harmon et al. 1997) that
such plateau states may correspond to major radio ejections from the
GRS 1915+105 system. Fig. 1 shows that a significant radio flare
occurred around MJD 50725. This was the prelude to a plateau state
lasting 20 days in which a bright, apparently optically thick, radio
source was formed, coinciding with a quasi-stable, hard X-ray
state. At the end of the plateau stage a second, major radio flare
occurred which triggered the MERLIN mapping observations, the highest
resolution imaging of GRS 1915+105 to date.  Each MERLIN map epoch is
indicated by a diamond in the upper panel of Fig. 1.

\section{Observations}

\subsection{MERLIN}

GRS 1915+105 was observed using the Multi Element Radio Linked
Interferometry Network (MERLIN) array at 4.994~GHz with a bandwidth of
16 MHz. MERLIN is comprised of six individual antennas typically 25m
in diameter and with a maximum separation of 217~km. Ten imaging runs
were made as a target of opportunity program triggered by flux
monitoring at the RT and GBI, showing the source had flared to around
250 mJy at 2--15 GHz. MERLIN, by default, samples both left hand and
right hand polarisations allowing the full range of Stokes parameters
to be imaged. Details of the observations can be found in the Table 1,
and continuum and polarisation maps in Figs. 2 and 6 respectively.

\begin{table*}
\begin{center}
\begin{tabular}{|ccccccccccccccc|cc|}
\hline
MJD     &    Total Flux   &  SE   &          &   M      & NW  &         &     M  &   C1 &          & C2 &           & C3  &  & \multicolumn{2}{c}{ Typical Errors}\\ 
--50000.0& Density & Posn. & Flux     &  P/I     & Posn. & Flux     &   P/I  &  Posn. & Flux     & Posn. & Flux      & Posn. & Flux & Flux & Posn. \\ 
\hline
752.71 & 179.8      & 46    &79.9      &  0.14  &15   &109.7
&   $<0.03$&     & & & & & & 0.4 & 6   \\      
753.71 & 139.9      & 69    &48.3      &  0.08   &23   &93.9&
$<0.03$&    & & & & & & 0.3 & 6   \\
754.70 & 85.2       &  99   &46.3      &  0.06   &31   &17.9
&   0.10& 38   &25.9&    &           &     &  & 0.3 & 6  \\
755.70 & 63.6       & 122   &31.0      &  0.06   &46   &11.7
&   0.13& 31   &21.4&    &           &     & &0.2 & 6 \\
758.68 &16.4       & 191   &4.8       &      &84   &4.8         &
& 92   &2.9 & 6 & 3.8&     & & 0.2 & 8  \\
759.68 &12.2       & 206   &2.0       &      &99   &3.1          &
& 130  &2.7&76  & 4.6&     && 0.1 & 10 \\
760.68 & 7.1      & 237   &1.8       &      &99   &2.4          &
&      &          &106 & 2.3&  &   & 0.1 & 10  \\
761.72 &7.4       & 252   &0.6      &      &107  &1.8&
& 206  &0.6&137 & 1.5&69   &1.6 & 0.1 & 15\\
762.72 &6.4       & 279   &0.4       &      &113  &1.4          &
& 211  &1.1&151 & 1.1&83   &1.1 &0.1 & 15\\     
763.72 &3.6       & 317   &0.3       &      &136 &0.9&        &
242  &0.4& 158 &0.5&98   &0.9 & 0.1 & 15\\
\hline
\end{tabular}
\end{center}
\noindent
\caption{Measured quantities from MERLIN maps: angular separation
(mas) from the core, flux densities (mJy), and fraction of linear
polarisation for all well-resolved components. SE and NW correspond to
the approaching and receding components of the first major
ejection. NW1 and NW2 correspond to the apparent splitting of
component NW in the later epochs (see text). C1 -- C3 are subsequent
approaching ejections. M = P/I is the fractional linear polarisation,
where measurable. The errors in fractional polarisation are typically
0.01 except for the last epoch quoted where they may reach 0.04 for
the weaker NW component.}
\label{details}
\end{table*}

Each separate epoch includes observations of a flux and polarisation
angle calibrator, 3C286, a point source calibrator, OQ208 or 0552+398,
the compact phase reference source, 1919+086 and the target
GRS~1915+105. By nodding the array to the phase reference source, the
interpolated phase variations due to atmospheric density variations
were removed from the target source data. The nodding cycle of 8
minutes on the target and 2 minutes on the reference source reflects
the phase stability of MERLIN at this frequency.

Phase referencing allows the registration of individual images to
around 10 mas, given the changeable weather conditions during the
observations. The absolute positions are tied to the radio reference
frame (ICRF) using the calibrator 1919+086 whose position is known to
around 17 mas relative to that frame (Patnaik et al 1992).

Using the MERLIN d--programs, initial data editing and amplitude
calibration were performed and the data prepared for further reduction
in the NRAO's AIPS package. After further editing in AIPS, each
dataset was run through the MERLIN pipeline, which images the phase
reference source and applies the derived corrections. Instrumental
polarisation corrections were made using 1919+086 and the polarisation
position angles were calibrated using 3C286, for which a position
angle of 33 degrees for the E vector of linear polarisation was
assumed.

Subsequent self-calibration using the phase reference images as
starting models was straight forward. Imaging the stokes parameters I,
Q and U using the AIPS task IMAGR gives the final total intensity
contour maps with superimposed vectors representing linearly polarised
intensity. All the images have been restored with a 40 mas (FWHM)
circular beam.

All the maps are affected to some degree by the variations in
source structure and brightness during the observations. In 12 hours, the
source expands by (typically) one-quarter of a beamwidth, and during
the first 4 epochs the flux-density changes were as much as 10\% over
each individual epoch.  Although these two effects are difficult to
disentangle, and we are in the process of simulating such changes to
quantify their magnitude, we believe that there will be no serious
consequences for the images beyond increased uncertainties in the
positions of components, at the level of a few mas, and the presence
of sidelobes around bright components (at a few percent of the peak
brightness) in the maps at early epochs.

Relative positions of components were found by measuring radial
distances of the components from a nominal map centre. RMS errors in
this process are estimated to be 6 mas, degrading to 15 mas for the
weaker components in the later epochs. The error in the absolute
position of each component is $\sim$ 22 mas.

A montage of all ten epochs is presented in Fig. 2. At least four
approaching (SE, C1, C2, C3) components and one receding (NW)
component are clearly resolved. The receding component corresponding
to ejection C1 may be also resolved in the last three epochs (less
likely, but possible, is a physical splitting of component NW).  The
maps have been rotated clockwise by 52 degrees, so that the mean
position angle of the jet on the sky is 142 degrees.

\subsection{Ryle Telescope}

The source was monitored at 15 GHz using the RT at
Cambridge.  At least a short observation was possible on most days
during this period.  The details of the observing technique are as
given in PF97. The data, in 5-minute bins, are
plotted in the top panel of Fig. 1. Data sampled at 32-sec intervals
are shown in Fig. 7 for the rapid (oscillation) variations observed at
four epochs.

\begin{figure}
\leavevmode\epsfig{file=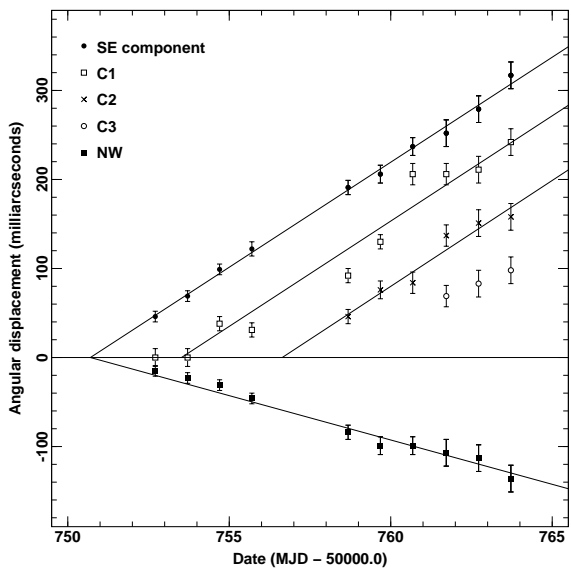,width=8cm,angle=0,clip}
\caption{Angular separation as a function of time for approaching
components SE, C1, C2 and C3 and receding component NW. All
approaching components are consistent with unaccelerated proper
motions of $23.6 \pm 0.5$ mas d$^{-1}$, significantly faster than
those reported in MR94. The receding component, NW, has a proper
motion of $10.0 \pm 0.5$ mas d$^{-1}$, and corresponds to the same
ejection epoch as SE, on MJD $50750.5^{+0.08}_{0.2}$.}
\end{figure}

\subsection{Green Bank Interferometer}

GRS1915+105 was monitored at 2.25 and 8.3 GHz throughout 1997 using
the 2-element Green Bank Interferometer.  The observing and
calibration procedures were the same as described by Foster et
al. (1996).  Random errors for the GBI are flux density dependent,
approximately (one sigma) 4 mJy at 2 GHz and 6 mJy at 8 GHz for fluxes
$<$ 100 mJy; 8 mJy at 2 GHz and 25 mJy at 8 GHz for fluxes near 0.5
Jy.  We estimate that systematic errors may approach 10\% at 2 GHz and
$>$20\% at 8 GHz.  Fig. 1 displays the GBI data at 2 and 8 GHz during
MJD 50700-50800 and exhibits both plateau (50730-50750) and flaring
(50721-50725 and 50750-50756) behavior as described by Foster et
al. (1996).

It should be noted that the {\em plateau} state discussed in this
paper is consistent with the definition in Foster et al. (1996) but
that our definition, i.e. flat spectrum radio emission at about 100
mJy at the same time as quasi-stable X-ray emission with a significant
hardening of the spectrum, may be more specific.

\subsection{RXTE}

The target is monitored up to several times daily in the 2-12 keV band
by the Rossi X-ray Timing Explorer (RXTE) All-Sky Monitor (ASM).
See e.g. Levine et al. (1996) for more details.  The total flux 
measured by individual scans is plotted in the bottom panels of
Figs. 1 and 7.

\section{Superluminal ejections}

Fig. 2 presents the ten MERLIN maps of GRS 1915+105.  The images
clearly show expansion of the source, with components on the left-hand
side (south east -- maps have been rotated clockwise by 52 degrees)
appearing to move faster.  The two components just resolved in the
first image can be followed through all epochs as they move out and
their brightness declines.  We label these components SE and
NW. Further components can be seen to be ejected later and are
labelled C1, C2 and C3. The proper motions of the components are
consistent with ballistic motions, and so we can extrapolate back to
derive their times of formation (see Fig 3). We estimate these to be
as follows (in MJD):

\[
{\rm NW + SE} : 50750.5^{+0.08}_{-0.2}
\]

\[
{\rm C1} : 50753.5 \pm 0.8
\]

\[
{\rm C2} : 50756.6 \pm 1.6
\]

\[
{\rm C3} : 50758.0 \pm 2.0
\]

\noindent
The latest time of the NW+SE ejection is constrained by the
observation of core-dominated radio oscillations by MJD 50750.58 (Fig.
7), as well as uncertainties in the model fits.  The individual
components are unresolved in the first 4 epochs.  There is some
evidence of them becoming resolved or breaking up into sub-components
in the latest 3 epochs.  Epochs 5, 6, 7 and 9 show that there is some
emission between the knots which may indicate the presence of an
underlying continuous jet. However, given the possible problems of
varying flux density and structure through the observations and the
low signal-to-noise ratio in the later epochs, we cannot be certain
about some of these features. The possible bending of the jet will be
discussed briefly below.

We will concentrate our quantitative analysis on those components
which can be followed through 3 or more epochs.  The rapidly outflowing
components are generally moving faster, and brighter, on the SE
side. (Note that the apparent high brightness of the NW component in
the second epoch is due to a blend with the
core).

From these components we can measure proper motions, flux densities
and (in some cases) polarisations. These we deal with quantitatively
below.

\begin{figure}
\leavevmode\epsfig{file=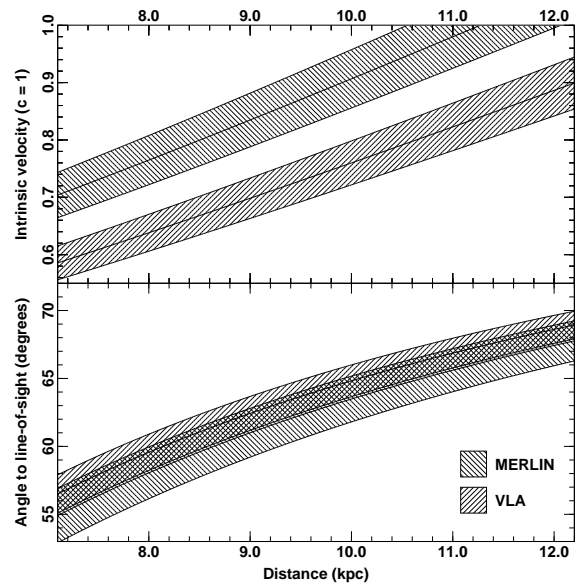,width=8cm,angle=0,clip}
\caption{Derived intrinsic velocities and
angles to the line-of-sight for MERLIN and VLA observations, under the
assumption of intrinsically symmetric ejections. Solutions
are possible for the same angle to the line-of-sight during both sets of
observations, but the velocity derived for the MERLIN
observations is {\bf higher} than that for the VLA observations, by
$\Delta \beta \sim 0.1$.}
\end{figure}

\subsection{Proper motions}

An examination of the positions of the components shows an apparent
 expansion velocity at greater than 2c, thus relativistic effects are
 important. As a result the approaching components will have a higher
 apparent motion than that of the receding components.  The component
 to the NW clearly has the lowest apparent speed we therefore identify
 it as receding; there are 3 components to the SE which we identify as
 approaching.  The data are best fit by proper motions of

\[
\mu_{\rm app} = 23.6 \pm 0.5 \phantom{00} {\rm mas} \phantom{0}{\rm d}^{-1}
\]

\noindent
and

\[
\mu_{\rm rec} = 10.0 \pm 0.5  \phantom{00} {\rm mas} \phantom{0}{\rm d}^{-1}
\]

\noindent
All fits are good, with $\chi^2_{\rm red} \leq 1$, and
illustrated in Fig. 3.  The proper motion of $17.6 \pm 0.4$ mas
d$^{-1}$ reported by MR94 for the approaching component can be ruled
out; fixing the proper motion to this value does not give an
acceptable fit to the data.

Following the method of MR94, {\em under the assumption of an
intrinsically symmetric ejection}, we can derive

\begin{equation}
\beta \cos \theta = \frac{\mu_{\rm app} - \mu_{\rm rec}}{\mu_{\rm app}
+ \mu_{\rm rec}} = 0.41 \pm 0.02,
\end{equation}

\noindent
where $\beta$ is the velocity of the ejections expressed as a fraction
of the speed of light, and $\theta$ is the angle between the ejection
and the line-of-sight. This immediately gives us a maximum angle to
the line-of-sight (setting $\beta = 1$) of $\theta_{\rm max} \leq
66 \pm 2^{\circ}$, and a minimum velocity (setting $\cos \theta = 1$) of
$\beta_{\rm min} \geq 0.41 \pm 0.02$.

A maximum distance to GRS 1915+105 can also be inferred for a maximum
possible velocity of the ejecta of $\beta = 1$

\begin{equation}
d_{\rm max} \leq \frac{c \tan \theta_{\rm max}}{2} \frac{(\mu_{\rm app} -
\mu_{\rm rec})}{\mu_{\rm app} \mu_{\rm rec}}
\end{equation}

\noindent
which can be expressed in convenient units as 

\begin{equation}
d_{\rm max} \leq 87 \tan \theta_{\rm max} \left( \frac{\mu_{\rm app}
- \mu_{\rm rec}}{\mu_{\rm app} \mu_{\rm rec}} \right) \phantom{00}{\rm kpc}
\end{equation}

\noindent
(for $\mu_{\rm app}$ and $\mu_{\rm rec}$ in mas d$^{-1}$).  The proper
motions observed with MERLIN therefore constrain the maximum distance
to be $11.2 \pm 0.8$ kpc.  MR94 state that GRS 1915+105 must be
further than the HII region G45.45+0.06 (mistyped by them as
G45.46+0.06; Downes et al. 1980) based upon their HI spectral
observations. While MR94 state that this HII region lies at $\sim 8.8$
kpc, in a recent study Feldt et al. (1998) adopt a distance of 6.6
kpc, and so GRS 1915+105 appears to lie at a distance of between 7 --
12 kpc, significantly closer than previously thought. Given the high
observed HI column density to the source, from which MR94 inferred a
large distance of $12.5 \pm 1.5$ kpc, we adopt a value of 11 kpc for
the distance to GRS 1915+105.

We can solve for the angle to line-of-sight and velocity, for any
distance to GRS 1915+105. The angle to the line-of-sight

\begin{equation}
\theta = \tan^{-1} \left[ 1.16 \times 10^{-2} \left( \frac{\mu_{\rm app}
\mu_{\rm rec}}{\mu_{\rm app} - \mu_{\rm rec}} \right) d \right],
\end{equation}

\noindent where $d$ is the distance to the source in kpc, and $\mu_{\rm app}$
and $\mu_{\rm rec}$ are in mas d$^{-1}$.  Once $\theta$ is calculated,
we can immediately derive $\beta$ as the value of $\beta \cos \theta$
is already known.

Table 2 lists the apparent velocities and solutions for the angle to
the line-of-sight and intrinsic velocities for both these MERLIN
observations and the VLA observations of MR94, for assumed distances
of 9, 10, 11 \& 12 kpc.  Fig. 4 shows these solutions for all
distances between 7 -- 12 kpc.  Our MERLIN data unequivocally imply a
higher intrinsic velocity, by $\Delta \beta \sim 0.1$, in comparison
to the observations reported in MR94.

\begin{table*}
\centering
\begin{tabular}{|ccccccccccc|}
\hline
Distance & \multicolumn{4}{c}{Apparent velocity} &
\multicolumn{2}{c}{Angle to} & \multicolumn{2}{c}{Intrinsic} &
\multicolumn{2}{c}{Bulk motion}\\
(kpc)    &  \multicolumn{2}{c}{(appr.)} & \multicolumn{2}{c}{(reced.)}
& \multicolumn{2}{c}{line-of-sight} & \multicolumn{2}{c}{velocity} &
\multicolumn{2}{c}{Lorentz factor}\\
\hline
        & MERLIN & VLA & MERLIN & VLA & MERLIN & VLA & MERLIN & VLA &
MERLIN & VLA\\
\hline
9         & $1.2c$ & $0.93c$ & $0.53c$ & $0.48c$ & $62^{\circ}$ &
$63^{\circ}$ & $0.84c$ & $0.71c$ & 1.8 & 1.4\\
10        & $1.4c$ & $1.1c$ & $0.59c$ & $0.55c$ & $64^{\circ}$ &
$66^{\circ}$  & $0.92c$ & $0.79c$ & 2.6 & 1.6\\
11        & $1.5c$ & $1.2c$ & $0.64c$ & $0.60c$ & $66^{\circ}$ &
$68^{\circ}$  & $0.98c$ & $0.86c$ & 5.0 & 2.0\\
12        & $1.5c$ & $1.2c$ & $0.66c$ & $0.63c$ & $68^{\circ}$ &
$69^{\circ}$  & $1.01c$ & $0.89c$ & -- & 2.2\\
\hline
\end{tabular}
\caption{Apparent velocities, solutions for the angle to the
line--of--site, intrinsic velocity of ejection (assuming symmetry),
and bulk motion Lorentz factors for values of proper motion measured
by MERLIN and previously by the VLA (MR94), for assumed distances of
9, 10, 11 and 12 kpc. The functions are plotted in full, with
consideration of measurement errors, in Fig. 4}
\end{table*}

\begin{figure}
\leavevmode\epsfig{file=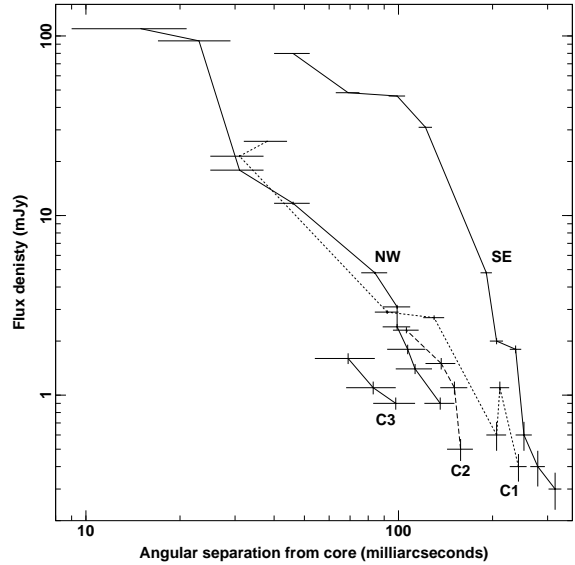,width=8cm,angle=0,clip}
\caption{Flux densities of individually-resolved components as a
function of angular separation from the core. 
Simple power-law and exponential decay models do not fit the
data well. The flux density ratio of SE : NW, approaching and receding
components from the same ejection, ranges from 6 -- 10.}
\end{figure}

\subsection{Flux densities and ratios}

The ratio of flux densities from the approaching and receding
components is another important diagnostic of the ejections.  In the
above calculations we have assumed geometric symmetry; here we also
assume that the jet components on both sides have the same intrinsic
luminosity (which does not appear to be the case for GRO J1655-40;
Hjellming \& Rupen 1995).  When measured at equal angular separations
from the core (i.e. the same time since ejection in the rest frame of
the ejecta) and coupled with the observed spectral index we can test
the theoretical predictions for Doppler (de-)boosting of the
components.

The mean spectral index of the ejecta is hard to determine due to both
overlapping ejection events and the presence of rapid flat-spectrum
oscillations from the core (PF97 and below). Our best estimate for the
index, defined as $\alpha = \Delta \log S_{\nu} / \Delta \log \nu$ is
around --0.8, the same as reported by MR94. This is determined
primarily from the simultaneous Green Bank monitoring at 13.3 and 3.6
cm and should be a good estimate for the 6 cm MERLIN observations.  It
is difficult to measure accurately the ratio of flux densities of the
components at equal angular separations, given their very different
proper motions, but it lies between 6 and 10. This is compatible with the
$8 \pm 1$ reported in MR94.

For bulk motions at velocity $\beta$ the observed ratio is
predicted to be 

\begin{equation}
\frac{S_{\rm app}}{S_{\rm rec}} = \left(\frac{1+\beta \cos
\theta}{1-\beta \cos \theta}\right)^{k-\alpha},
\end{equation}

\noindent where, theoretically, $k = 2$ for continuous jets and 3 for
discrete components. As $\beta \cos \theta$ is already calculated as
$0.41 \pm 0.02$, and independent of distance, we can solve for
$k$. For a flux density ratio of in the range 6 -- 10, $k$ = 1.3 --
1.9. As in MR94, the flux ratio appears to be closer to that expected
for a continuous jet than for discrete components (although the value
derived from the data of MR94 is $k = 2.3$, interpreted by them as
implying something intermediate between a continuous jet and discrete
ejections). For $k=3$ we would have expected a flux ratio of $\sim 27$.

Fig. 5 shows the flux density of each well-observed component as a
function of angular separation from the core. As is obvious from Table
1 and the maps of Fig. 2, the ejections in the sequence NW+SE, C1, C2,
C3 are of steadily decreasing flux density. While uncertainties in
both flux density and position make interpretation difficult, the data
do not appear to show the straight line behaviour expected for a
power-law decay (for the case of adiabatic expansion losses and a
constant expansion rate). Exponential fits to the data provide no
improvement.  RM99 discuss a steepening of the decay of the radio flux
density at an angular separation of $\sim 1$ arcsec; as we only image
on smaller scales we cannot test this. Atoyan \& Aharonian (1997)
discuss the implications of the flux ratio in considerable detail.

\subsection{Polarisation}

Polarisation images have been made for the first four epochs, when the
source was sufficiently bright to detect linear polarisation at the
level of a few percent.  These are shown in Fig. 6, where the vectors
represent the fractional linear polarisation and observed electric
field angle. The fractional linear polarisations are also listed in
Table 1.  No circular polarisation was detected at any time, with a
conservative upper limit of 2\%.

\subsubsection{Resolved ejecta}

The images of the ejected components show a striking
asymmetry in the linear polarisation -- only the approaching SE
component appears significantly polarized. Its fractional polarisation
decreases rapidly from 14\% to 6\%, and the polarisation position
angle swings by approximately 75 degrees between the second and third
epochs, then swings back by 45 degrees.  In the fourth epoch image,
there is marginal detection of polarized emission in the receding NW
component, at a level of $13 \pm 5$\%.

The changes in polarisation seen in the MERLIN images could be a
result of changing Faraday effects (internal or external) or changing
magnetic field geometry within the radio components.  With only a
single frequency it is hard to distinguish between these possibilities.
If the observed change in position angle is due to Faraday rotation,
the implied change in rotation measure is $> 300$ rad m$^{-2}$.

\begin{figure*}
\leavevmode\epsfig{file=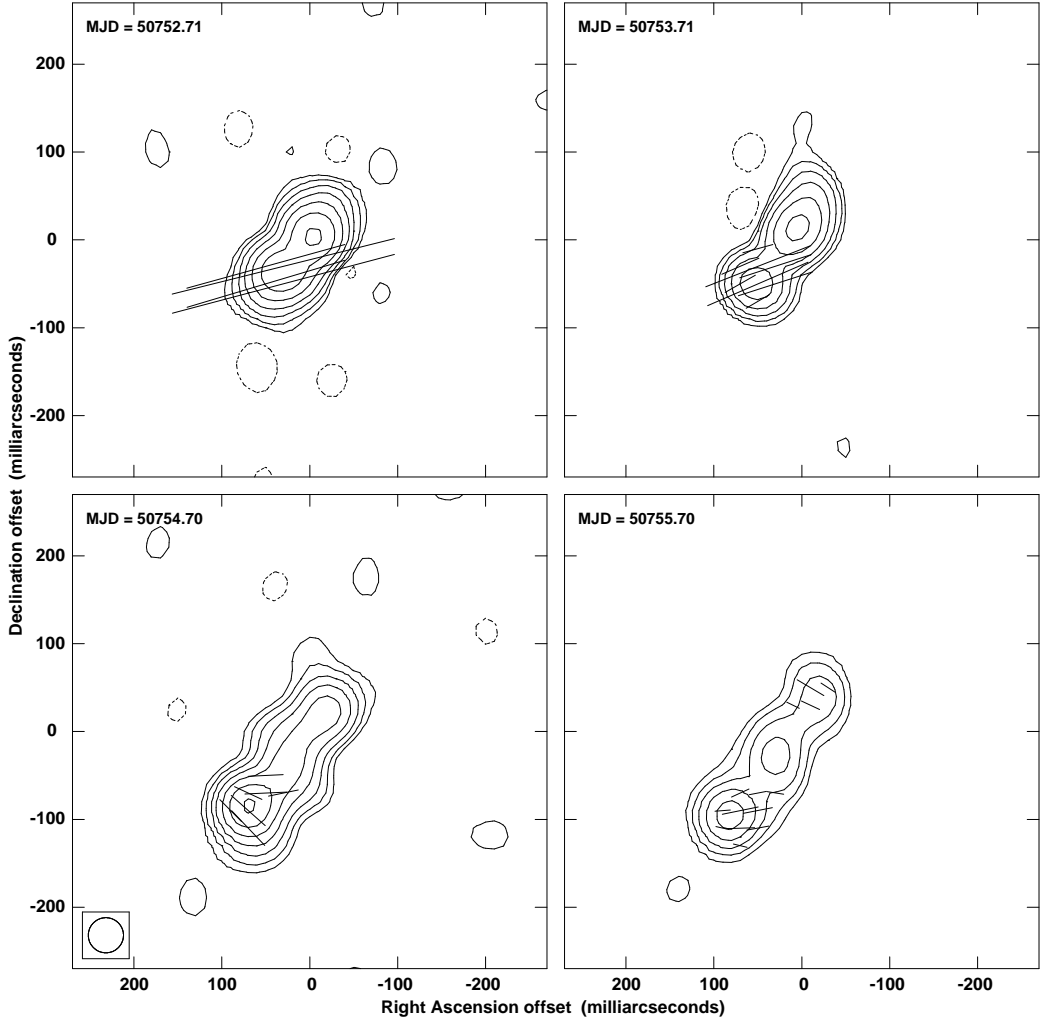,width=14cm,angle=0,clip}
\caption{Linear polarisation E vectors superimposed upon total
intensity contour maps of the first four epochs of MERLIN
observations. For polarisation vectors, 100 mas = 3.33 mJy/beam. This
clearly shows the depolarisation of the core and receding components
with respect to the approaching component, and the rotation of
polarisation vectors in the approaching component.}
\end{figure*}

The radio components are believed to be considerably denser than their
surroundings (observed limits on the deceleration imply a density
contrast of at least 10) and have equipartition field strengths of order
50 mG (see below), presumably much greater than in the surrounding
medium.  The Faraday depth, or rotation measure, is proportional to
$nB(Ld)^{1/2}$, where $n$ is the thermal electron density, $B$ is the
magnetic field strength, $L$ is the path length and $d$ is the field
tangling scale. The Faraday depth within the source is therefore
likely to exceed the Faraday depth of a surrounding medium distributed
on a scale size comparable to the source itself.  As the components
expand their internal Faraday depths will fall, and any internal
Faraday depolarisation would decrease. This is contrary to the
observed decrease in fractional polarisation.  Therefore, we do not
believe the changes in polarisation are due to changing Faraday
depths, and this is consistent with the low rotation measure observed
by Rodr\'\i guez et al. (1995).

Rodr\'\i guez et al. (1995) noted a fractional polarisation of
approximately 2\% at 5.0, 8.4 and 15 GHz at a position angle of $178
\pm 10$ degrees on 24 March 1994, when the source was barely resolved
by the VLA and 5 days after the radio components were ejected.  At
this time the emission was probably dominated by the approaching
component.  The similarity of the fractional polarisation and position
angle at the three wavelengths implies that Faraday effects were small
at 5.0 GHz.  The implied upper limit to the Faraday rotation measure
is 50 rad m$^{-2}$.

Superluminal jets in AGN often show complex radio polarisation
behaviour: rapid variations in polarisation position angle (e.g.
0917+624 for an extreme case, Quirrenbach et al. 1989) and
small-scale variations along the jet (e.g. 3C454.3, Kemball, Diamond
\& Pauliny-Toth 1996).  This is usually interpreted as being due to
shocks, which compress and order the magnetic field (Cawthorne et
al. 1993), and which appear as radio knots in an otherwise continuous
jets.

The radio components of GRS1915+105, on the other hand, are thought to be
dicrete plasmons, rather than features (shocks or otherwise)
in a continuous jet.  In this case the changes in radio polarisation
indicate changes in the internal magnetic field geometry, such as
randomization of the magnetic field as the plasmons evolve.
The effects of aberration on a planar structure such as a shock would
result in the two components having different observed polarisation
characteristics. However the low signal to noise for the measurement
of polarisation in the receding component does not allow us to make
such a distinction in this case.

\subsubsection{Core}

The core, sometimes blended in our images with newly formed major
ejections, does not at any time show significant linear
polarisation. In particular, during the first two epochs of
observation, the core is dominated by the flat spectrum oscillations
of the type discussed in detail in PF97 (see Fig. 7). The lack
of observed polarisation may be due to superposition of multiple
components with different polarisation position angles within the
beam, or large Faraday depths close to the binary system (either
internal to the compact base of the jet or in the form of
circumstellar material).

\subsection{An underlying curved jet ?}

In several of the MERLIN maps there is a hint of extended emission
between the components. Furthermore, the low value of $k$ derived from
the flux density ratios and $\beta \cos \theta$ is more consistent
with a continuous jet than physically discrete components (see
above). In addition, the jet appears to be curved, an effect which may
also be marginally apparent in reinspection of the VLA images of MR94.

Firstly, comparing the flux in discrete components in the MERLIN maps
with respect to radio monitoring with GBI and RT (in which the source
and jets will be unresolved), we find that the values agree to within
a few per cent. So the overwhelming majority of the synchrotron
emission, at cm wavelengths at least, arises in unresolved (with
MERLIN) bright spots. On the other hand, we both derive (section 3.2)
and apply (section 4.1) $k \sim 2$, the value theoretically expected
for a continuous jet. We must conclude that the data are not good
enough at present to definitively establish whether the ejections are
in discrete components or we are observing the bright parts of a
continuous jet.  In the latter case we can at least now state that any
underlying continuous jet is very faint and less than a few mJy at cm
wavelengths

The possible curvature of the jet is not addressed quantitatively
here, but we do believe this effect to be real. Discrimination between
ballistic and helical effects, the latter possibly due to
Kelvin-Helmholtz instabilities, is not possible at this stage. The
`period' of any bending observed at angular scales imaged with MERLIN
is around 7 days. It is interesting to note that the `period' of the
jet bending in GRO J1655--40 reported by Hjellming \& Rupen (1995) of
$3.0 \pm 0.2$ d turned out to be very close to the subsequently
discovered orbital period of 2.6 d (Bailyn et al. 1995).  We are
confident that the apparent bending is not due to the rapid angular
and flux density evolution of the source duing a 12-hr observation
(c.f. however, VLBI observations of GRO J1655--40, Tingay et al. 1995,
where both higher proper motions and finer angular resolution
exacerbated the effect).

\subsection{Collimation and expansion of components}

From the lack of clearly resolved structure perpendicular to the jet
axis at the 40 mas resolution of MERLIN, we can place constraints on
the opening angle and lateral expansion of the ejected components. The
maximum distance from the core that we track ejected components is
$\sim300$ mas, and this constrains the opening angle of the jet to
$\leq 8$ degrees over distances from the core $\leq 4000$ A.U. ($6
\times 10^{16}$ cm). Similarly the ratio of lateral expansion to
forward velocities is constrained to be $\leq 0.14$, i.e. a maximum
lateral expansion velocity $\leq 0.14$c (for jet bulk velocity of
$c$).

\section{Discussion}

\subsection{Energetics and mass flow}

\subsubsection{Estimation of the internal energies of the jet components}

Initially, we must work in the rest-frame of the emission region.
Converting the observed parameters (emission frequency, luminosity etc)
requires the value of the bulk
Lorentz factor $\gamma$, which is somewhat uncertain as discussed in
section 3.1, on account of the uncertainty in the distance.
Because of the large angle to the line-of-sight, the Doppler factors
for both the approaching and receding components,

\begin{equation}
\delta_{\rm app, rec} = \gamma^{-1} (1 \mp \beta \cos \theta)^{-1}
\end{equation}

\noindent are both less than unity for the range of Lorentz factors
considered here.

The estimation of the parameters of the synchrotron emission region is
detailed by Longair (1994) and Hughes (1991). We adopt the formulae in
the summary by Longair (allowing for the opposite convention for the
sign of $\alpha$). These include a number of simplifying assumptions
which are unlikely to introduce larger uncertainties than those
imposed by our imperfect knowledge of the source parameters.

In order to estimate the internal energy of the emitting component,
we need to know the synchrotron spectrum and the source geometry.
A radio spectrum of the form $ S \propto \nu^{\alpha} $ arises from an
electron population with an energy spectrum \( N(E) \propto E^{-(1 - 2
\alpha)}\).  The observed luminosity can arise from various
combinations of magnetic flux density {\it B} and particle density,
both of which we assume to be spatially uniform; the total energy is
minimised to derive the ``minimum energy'' conditions: the total energy
in relativistic electrons, and the misleadingly-named $B_{\rm min}$,
the field at which the total energy is minimised.  As is well known,
the minimum energy result nearly coincides with equal energies in
magnetic field and particles, or equipartition. There are physical
plausibility arguments for such a situation, but very few measurements
which confirm that this situation might actually pertain in any
particular case (see Harris, Carilli \& Perley 1994 for data on the
radio galaxy Cygnus A which suggest, in that case, that conditions may
be near to those suggested by this method).

\begin{table*}
\caption{A summary of the derived properties of the ejecta at minimum energy
conditions, calculated for a distance of 11 kpc, and summing both
sides of the ejection.
We assume that radio
emission observed by GBI and RT at the peak of the flare are dominated by
the approaching (SE) component, and take 1 and 15 GHz as upper and lower
bounds of the observed flaring emission respectively. We assume a
generation time for the ejection of 12 hr. Proton mass and mass flow
rate are for the case of 1 proton for each electron.}
\centering
\begin{tabular}{cccccccc}
\hline
\multicolumn{3}{|c|}{Minimum energy condition} &
\multicolumn{3}{|c|}{Ejecta} & Power & Mass outlflow \\
E$_{\rm min}$ (erg) & E$_{\rm K, min}$ & B$_{\rm min}$ (mG) &
N$_e$ & M$_e$(rest) (g) & M$_p$ (g) & (erg s$^{-1}$)  & rate (g
s$^{-1}$)  \\
\hline
$2 \times 10^{43}$ & $2 \times 10^{44}$ & 280 & $3 \times 10^{46}$ & $3 \times
10^{19}$ & $5 \times 10^{22}$ & $2 \times 10^{39}$ & $10^{18}$ \\
\hline 
\end{tabular}
\end{table*}

The calculation can easily be adapted to include
a contribution to the energy by a hypothetical population
of relativistic protons, which do not radiate significantly by the
synchrotron mechanism if their energies are comparable with those of
the electrons.

If the energy spectrum of the electrons is relatively steep
$(\alpha$ steeper than  $-0.5)$, as in this case,
the total energy of the particles is dominated by the lower end of the
distribution.  The lowest frequency detected, and the luminosity at that
frequency, are then parameters in the calculations.

The source geometry is, so far, not well-defined. Direct observations
only give an upper limit to most dimensions: the resolution of MERLIN
is 50 mas, corresponding to $8 \times 10^{13}\,\rm{m}$, or 3 light
days, at 11 kpc.  We have to rely on the time-scale of the variations
in flux density for a better limit; the rise times are the shorter and
therefore more restrictive. The rise time may represent the interval
during which relativistic material is injected, or alternatively a
transition in optical depth (or both). We believe that the former
(injection timescale) is more probably the dominant effect, since the
spectral data suggest that the source is optically thin as soon as the
jet is visible.

Since the rise-time for the jet components appears to be less than 12h,
we adopt 12 light-h $(1.3 \times 10^{13}\,\rm{m})$ as a ``typical'' size on
formation of the jet component.
In comparison, RM99 discuss a geometric mean angular size
of 35 mas, corresponding to a linear scale of $6 \times 10^{13}\,\rm{m}$.
We stress the weak links in this argument, elaborated by Longair (1994)
and by Hughes (1991):
\begin{itemize}

\item We have only rather weak constraints on the source size, and
certainly do not know the filling factor of the source, and this
is one of the major uncertainties in the calculation. A smaller size
or filling factor reduces the total energy required as $({\rm volume})^{3/7}$.

\item There is little evidence that the minimum-energy condition is achieved.

\item The contribution of protons (and other nuclei) is not known.

\item The limits on the energy spectrum are not well-defined by current
observations.

\end{itemize}

We now apply the formulae from Longair (1994); his equations 19.29 and 19.30,
using the following measured, estimated or derived parameters:

\begin{enumerate}

\item adopted distance 11 kpc

\item $\gamma = 5.0$ (see Table 2)

\item $\beta \cos \theta = 0.41$

\item spectral index $\alpha = -0.8$

\item Doppler factors $\delta_{\rm app, rec} = \gamma^{-1} (1 \mp \beta \cos \theta)^{-1} = 0.34, 0.14 $

\item $k = 2$ (section 3.2)

\item ratio of relativistic proton energy to electron energy, $(\eta - 1) = 0 $

\item total source volume $V = 10^{39} \, {\rm m}^{3}$

\item monochromatic luminosity $L_{\nu}$ in the rest frame, at the lowest
frequency detected, as derived below.

\end{enumerate}

The maximum observed flux density observed at the start of the
outburst at 2.3\,GHz is 550 mJy (Fig 7), some 2 days before the first
MERLIN map. This would correspond to 1\,Jy at 1\,GHz, assuming the
spectral index to be $-0.8$. We adopt 1\,GHz as the lowest frequency
at which emission was detected during this outburst (Hannikainen \&
Hunstead, private communication, detected the outburst at 843\,MHz).
We assume that this flux density is dominated by the approaching
component.  Its apparent flux density if observed in its rest frame
would be about 20 Jy; signals received at 1\,GHz would have been
emitted at $\nu =$ 3\,GHz, and the luminosity $L_{\nu}= 2.9 \times
10^{17} \,{\rm W\,Hz}^{-1}$.

The minimum total energy  is

\begin{equation}
W_{\rm min} \simeq 3.0 \times 10^{6}\, \eta^{4/7} (V/{\rm m^{3}})^{3/7}
(\nu/{\rm Hz})^{2/7} (L_{\nu}/{\rm W\, Hz^{-1}})^{4/7} \, {\rm J}
\end{equation}

\noindent and the associated magnetic field is

\begin{equation}
B_{\rm min} \simeq 1.8\, (\eta (L_{\nu}/{\rm W \, Hz^{-1}})/(V/{\rm m^{3}}))^{2/7} (\nu/{\rm Hz})^{1/7} \, {\rm T}
\end{equation}

This set of parameters, for one of the radio-emitting components, leads to
$W_{\rm min} = 7.7 \times 10^{35}\,{\rm J} = 7.7 \times 10^{42}\, {\rm erg} $
and $B_{\min} = 28\, \mu {\rm T} = 280\, {\rm mG}$.

The Lorentz factor of an electron radiating near 3\,GHz is then about
90, and the mean Lorentz factor for this population is 240.  (Some
authors consider electron distributions which continue to much lower
energies, but as yet there is no observational evidence for this.
RM99, on the other hand, adopt a higher mean Lorentz factor of 1000, on
the basis of a detection of the event at 240\,GHz.  That emission may
well have originated in the inner region of the system; Fig 7 shows
the quasi-periodic oscillations observed at 15\,GHz during the early
stage of the current outburst which have a flat spectrum (Fender \&
Pooley 1998) and, with a time-scale of 20\,min, must come from a very
small region.)

\subsubsection{Total number of relativistic electrons}
By integrating the electron distribution we can derive an estimate
of the total number of relativistic electrons (this may become important
if we suppose that each is accompanied by a ``cold'' proton).
This integration requires minimum and maximum energies, but as in the
derivation of the total energy the precise value of the upper limit is
unimportant; the number is dominated by the low-energy electrons.
To establish the constants involved we use equation 19.17 from Longair,
which relates the luminosity, the magnetic field and the spectrum,
and derive
\[
N_{\rm total} = (L_{\nu}/{\rm W\, Hz^{-1}})f(\alpha)/(B_{\rm min}/{\rm T})
\]
\noindent where $f(\alpha) = 1.7 \times 10^{24}$ for $\alpha = -0.8$\,.
This results in $N_{\rm total} = 1.6 \times 10^{46}$.

%
%
%

\subsubsection{Kinetic energy}

Associated with the bulk motion of the jet there is also the kinetic
energy; if the jet contains only $e^+e^-$ plasma and magnetic field,
the magnitude of this is $(\gamma - 1) \times W_{\rm min}$,
perhaps $9 \times 10^{36}\,{\rm J} = 9 \times 10^{43}\, {\rm erg}$.
On the other hand,
if the plasma contains cold protons, the kinetic energy will be
increased by $(\gamma -1) \times N_{\rm total}\, m_{\rm p}\, c^2$.
This results in an increase of only a factor 2.
The total of the kinetic energy of the protons could
vary, either way, if the conditions were far from ``minimum energy'',
or be substantially larger if the electron energy distribution continues
to rise to lower energies, with correspondingly more protons.

We can estimate the power required to produce this injection of energy
over 12\,h; including the kinetic energy of the system and both sides
of the source, we have to generate $2 \times \gamma W_{\rm min}$ in
that time, a power of $2 \times 10^{32}\, {\rm W}$ or $2 \times
10^{39}{\rm erg\,s^{-1}}$. Including a similar number of protons
approximately doubles that total and also requires a mass-flow rate of
$\geq 10^{18} \rm g\,s^{-1}$. X-ray spectral fits (e.g. Belloni et
al. 1997b) have been used to derive accretion rates near $10^{18} \rm
g\,s^{-1}$.

Many of the observed parameters used here are uncertain, with the dominant
effect arising from the unknown distance. Were we to adopt a distance of 9 kpc,
the bulk Lorentz factor would be 1.8 rather than 5 and the Doppler factor
of the approaching component near 1. Therefore its flux density in the rest
frame would be 1\,Jy rather than 20\,Jy, and the energy of the system
reduced by a factor of about 10. 

The results of these calculations of energetics and mass flow are
summarised in table 3 (for a distance of 11 kpc). Indeed the mass of
cold protons could be as much again as in table 3 if we use the upper
limit on electron density ($<2$ cm$^{-3}$) inferred from Faraday
effects discussed in section 3.3.1.

\subsection{Comparison with previous major ejections}

While it could already be guessed from the higher observed proper
motions, our solutions (Table 2, Fig. 4, under the assumption of intrinsic
symmetry) illustrate that we cannot avoid deriving a higher velocity
for the ejecta than derived by MR94.  Thus the two data sets are {\em
not} compatible with a simple change of the angle to the line-of-sight
of a jet which is intrinsically physically identical. In fact the data
{\em are} marginally consistent with the {\em same} angle to the line
of sight for the jet. The increase in intrinsic $\beta$ from our
observations compared to MR94 is at least 0.05, and more likely $\sim
0.1$, corresponding to a significantly higher Lorentz factor for bulk
motion by a factor of $\geq 1.3$, as tabulated in Table 2.

It is of great importance to the determine whether the different
measured velocities are intrinsic to the ejecta or an artefact of the
different resolutions of MERLIN and the VLA. In the former case the
significantly different measured proper motions could correspond to a
difference between the ejection event(s) reported here and
the event reported in MR94. The earlier event appears to have been
both significantly brighter and to have decayed in flux more slowly.
Alternatively, there may be genuine deceleration between the angular
scales of up to 300 mas imaged with MERLIN and greater than 400 mas
measured with the VLA. This does not seem likely however, given that
both instruments recorded essentially ballistic motions (ignoring the
slight apparent curvature) within their multiple observations.

It may be that the different measured proper motions arise from an
inability of the VLA to resolve individual components, causing them to
blend together in maps. In this case, the VLA could measure lower
proper motions and/or apparent decelerations if 

\begin{enumerate}
\item[(1)]{There were multiple components, and}
\item[(2)]{The components decreased in flux density more rapidly with
distance from the core.}
\end{enumerate}

This effect is discussed by Hjellming \& Rupen (1995) for GRO J1655--40
where they measure apparent decelerations of $\sim 30$\% with the VLA
in comparison to VLBA (and SHEVE -- Tingay et al. 1995) for two of the
three major ejections that they image.  Did these effects occur during
the observations of MR94 ? Their maps do indeed show that multiple
ejections were occurring during this period (MR94; RM99); and
furthermore RM99 discuss evidence that the decay rate of the ejecta
increase with angular separation from the core (although they discuss
a relatively abrupt increase in decay rate at angular separations
$\geq 1$ arcsec, which would not have affected the measurements of
MR94). So it seems a possibility that the lower measured proper
motions arise from blending of multiple components and that the MERLIN
measurements more accurately represent the true situation.

In order to test this we have attempted to convolve our MERLIN data
with lower-resolution beams comparable to that of the VLA observations
of MR94. Note that the comparison is not good because of the lack of
short baselines in the MERLIN array compared to the VLA. Nevertheless,
it was clear from convolution of our data with a 250 mas beam that it
was not possible to clearly resolve any individual components. Even
with a 100 mas beam, i.e. twice the resolution of the VLA data of
MR94, while the approaching and receding sides can be distinguished,
individual ejecta on the approaching side cannot.

However, without simultaneous VLA and MERLIN (and possibly also VLBI)
observations of the outburst, it is not possible to test this
conclusively. So we cannot confidently discriminate between the two
most likely explanations, of either intrinsically different velocities
between March/April 1994 and Oct/Nov 1997, or resolution effects
between the two arrays. Note that if the latter is the cause of the
different measured proper motions then it requires the similar proper
motions (range 15 -- 18 mas d$^{-1}$) measured in four instances in
1994 with the VLA (RM99) all to have an origin in similar unresolved
multiple ejections. Detailed inspection of radio light curves around
these periods may shed some light on this.

The mean position angle of the jet as observed by MERLIN, 142 degrees,
is consistent with the range of 140 -- 160 degrees observed with the
VLA (RM99).

\begin{figure*}
\leavevmode\epsfig{file=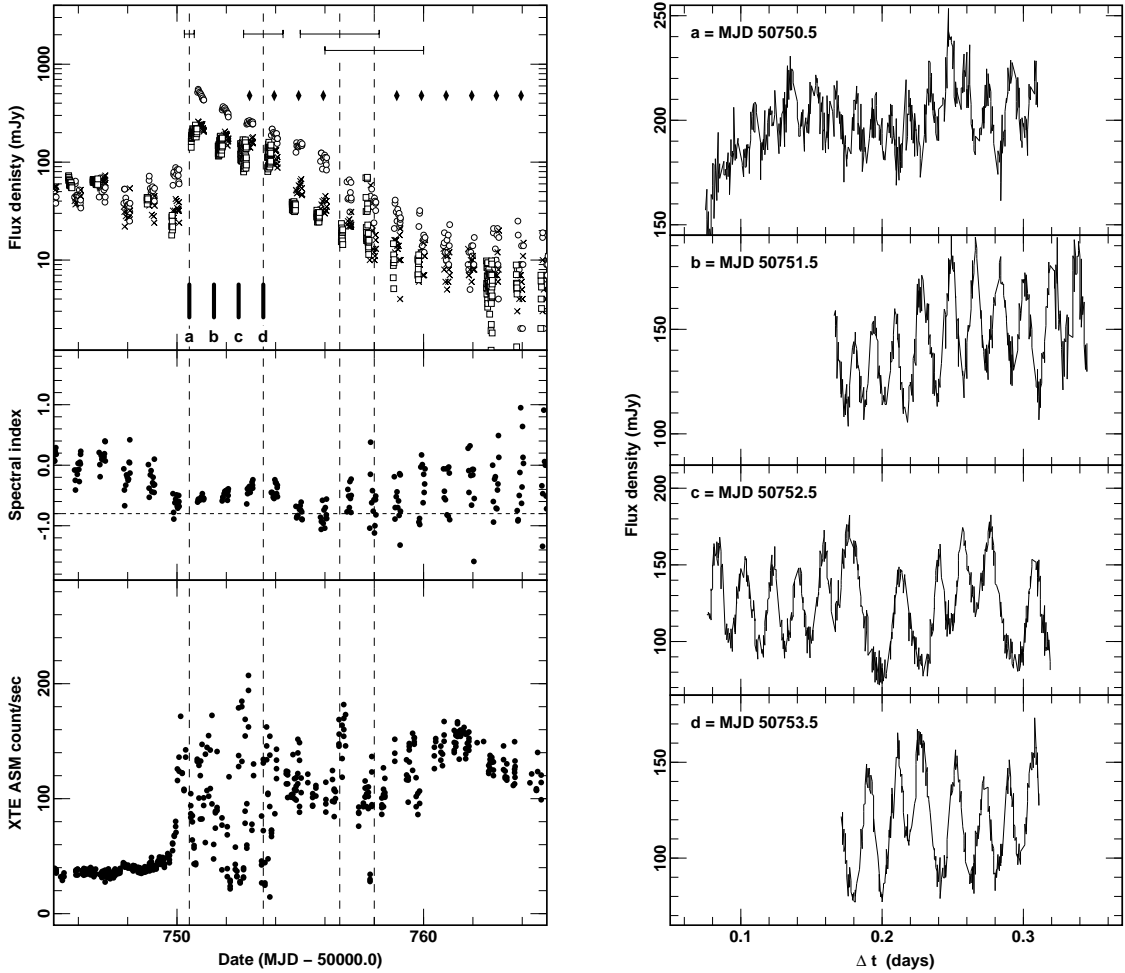,width=16cm,angle=0,clip}
\caption{A detailed view of the radio and X-ray behaviour of GRS
1915+105 during the ejection episode. Symbols are as shown in Fig. 1.
The vertical dotted lines, with error bars, indicate our derived times
for the beginning of each of the four ejections.The four solid
vertical bars labelled {\bf a -- d} drawn on the top left radio monitoring
panel between the first and second ejections, indicate the four periods of
15-GHz monitoring highlighted in the panels on the right.  The source
appears to exhibit continuous short period (20--40 min) oscillations
in this phase, indicative of repeated advection or ejection and refill
of the inner accretion disc.  The radio QPO do not repeat after the
second ejection, but there are some indicators of the second and third
ejections in the X-ray data.}
\end{figure*}

\subsection{X-ray and radio state at jet formation}

GRS 1915+105 offers a unique opportunity to study the jet -- disc
relation, and in this case we can directly relate changes in the
accretion disc X-ray emission to the ejection of significant
quantities of material from the system.

Fig. 1 shows the evolving sequence of events as observed by radio and
X-ray monitoring of the source. The onset of the plateau phase is
preceded by a significant radio flare, similar enough to those mapped
by us after the plateau state to interpret it as a major disc
ejection. 

Almost immediately following this ejection the plateau state is
established. It was already known from PF97 that
this state, with hard, quasi-steady X-ray emission, corresponded to bright
and fairly steady radio emission at 15 GHz. Combination of the RT
and GBI data now clearly show that this state is indeed radio bright,
with a flat or inverted spectral index. This is highly indicative of
absorbed emission (although it is conceivable that the electron
acceleration mechanism changes to produce a different, optically thin,
spectrum which mimics absorption), and may suggest the formation of a
large optically thick jet. If so, it seems possible that this may be
associated with the infrared jet imaged by Sams, Eckart \& Sunyaev
(1996). The X-ray emission may also support such a picture, as the
soft (disc) component appears to be very weak, if present at all,
during this state (Mendez \& Belloni, private communication). In
addition, no radio oscillations, associated with inner disc
instabilities, have been observed during the plateau state.

Fig. 7 shows in more detail the radio and X-ray state of the source at
the times of formation of the jets we have imaged with MERLIN.  In
particular note the four panels on the right hand side of Fig. 7,
showing the 15--GHz RT monitoring of GRS 1915+105 between MJD 50750
and 50754, i.e. between the first ejection, corresponding to NW+SE,
and the second, corresponding to C1. It appears that for the entire
period between these two ejections, the inner accretion disc is
unstable and material is repeatedly advected and ejected on timescales
of tens of minutes (see Belloni et al. 1997a,b; Fender et al. 1997;
PF97; Eikenberry et al. 1998; Mirabel et al. 1998; Fender \& Pooley
1998).  No such periods of large amplitude inner disc instabilities
are obvious between subsequent ejections, although the ejection of C2
appears to correspond to a minor X-ray flare, and C3 possibly
corresponds to a brief X-ray dip.

\section{Conclusions}  

Our MERLIN observations have revealed four major relativistic
ejections from GRS 1915+105 over a period of approximately two
weeks. Over ten epochs of observation we have consistently measured
proper motions of $23.6 \pm 0.5$ and $10.0 \pm 0.5$ mas d$^{-1}$, for
the approaching and receding components respectively. The proper
motion of the approaching component is more than 30\% higher than that
reported from VLA observations of the source (MR94; RM99). Under the
assumption of intrinsic symmetry, we have derived an upper limit for
the distance to the system of $11.2 \pm 0.8$ kpc. While compatible
with the distance of $12.5 \pm 1.5$ kpc quoted by MR94, it seems that
the favoured distance to the system should be revised downwards by at
least 1 kpc. Solving for angle to the line-of-sight and intrinsic
velocity, again under the assumption of an intrinsically symmetric
ejection, we cannot avoid deriving a significantly higher velocity
than MR94, by around $\Delta \beta = 0.1$.  We have investigated
whether or not the lower resolution of the VLA could result in lower
proper motions being measured, as a result of multiple blended
components and an increase in decay rate with distance from the core,
and this does seem possible. On the other hand, there is no reason to
believe that the jet velocity should be fixed -- of all the
astrophysical objects with relativistic jets, only SS 433 (jet
velocity 0.26$c$, e.g. Vermeulen et al. 1993) is established to have a
constant jet velocity.  Further observations at high angular
resolutions with MERLIN and/or VLBI, preferably simultaneous with
lower resolution VLA observations, will be required to investigate
whether there is a systematic deceleration of ejecta on angular scales
of $\geq 0.3$ arcsec.  If MR94 are correct in inferring a large
distance for GRS 1915+105 from the column density to the source, then
the bulk velocity of the outflow is almost certainly much higher than
the currently accepted value of 0.92 $c$; at 11 kpc we derive
$0.98^{+0.02}_{-0.05}c$ at $66 \pm 2$ degrees to the line-of-sight.

Our polarisation observations clearly reveal rapid evolution of the
magnetic field in the ejecta on timescales of a day or less. This
would seem to imply that the region from which the polarised emission
arises is smaller than one light day across; consideration of time
dilation (stretching intrinsic timescales when observed in our frame,
given our derived solutions for $\theta$ and $\beta$) only makes this
size smaller. The decreasing trend of the polarisation with distance
from the core suggests increasing randomisation of the field as the
ejecta evolve, although multi-frequency polarisation measurements are
required to rule out or constrain Faraday rotation effects.

As already noted by MR94 and others, the power required for the
formation of the jet is immense, far greater than the Eddington
luminosity for a solar mass object (at 11 kpc), even without the
inclusion of a proton content.  Jet formation may well be the dominant
power output channel during such periods, and possibly also during
periods of smaller oscillation ejections. In addition, when one proton
per electron is added the minimum mass flow rate becomes comparable to
the mass accretion rates derived from X-ray spectral fits. So, it is
possible that during jet formation periods a significant fraction of
the inflowing mass is expelled and does not fall into the black
hole. A similar possibility exists for the minor oscillation ejections
(Fender \& Pooley 1998). Good wavelength coverage of flares to both
higher and lower frequencies is required to better determine the
luminosity, energy and mass of the ejections. An accurate simultaneous
comparison of mass flow through the disc (from X-ray spectral fits)
and outflow rates (from radio observations) would be of great
interest, to see whether the flows are advection-- or
ejection--dominated.

Radio and X-ray monitoring of GRS 1915+105 (e.g. Harmon et al. 1997;
PF97) had already hinted at a delayed relation between hard X-ray
states and radio outbursts. These MERLIN observations have established
that these radio outbursts do indeed correspond to relativistic
ejections following plateau states. The nature of these states, and
the associated inverted spectrum radio emission is still unclear, and
warrants further study. It is also of interest to clarify whether or
not the radio flare which {\em precedes} the plateau state (Fig. 1)
also, as expected, corresponds to a major ejection. Why the accretion
disc -- jet system in GRS 1915+105 appears to switch so rapidly
between major mass ejections, short period (oscillation) instabilities
and back again (Fig. 7) is still very uncertain. Atoyan \& Aharonian
(1997) discuss the disruption of the accretion disc due to recoil
momentum from an asymmetric ejection. However, it is difficult to
reconcile their model with the apparent very rapid reformation of the
(unstable) inner accretion disc after the major (NE+SW) ejection.
Finally, it is tempting to ascribe the apparent `period' of the jet
bending of $\sim 7$ days to an orbital period, as was found to be the
case for GRO J1655--40, but at present this is no more than
speculation.

These MERLIN observations are further evidence that GRS 1915+105
repeatedly produces relativistic ejections of massive
clouds of synchrotron--emitting electrons. Our observations of very
high proper motions in the inner 0.3 arcsec of the jet show that
galactic stellar--mass black holes are capable of accelerating matter
to velocities very close to the speed of light. 

\section*{Acknowledgements}

We acknowledge the assistance of many people in the triggering and
realisation of these observations, including Shane McKie, Richard
Ogley and Peter Thomasson.  MERLIN is operated as a National Facility
by the University of Manchester at the Nuffield Radio Astronomy
Laboratories, Jodrell Bank, on behalf of the Particle Physics and
Astronomy Research Council (PPARC). We thank the staff at MRAO for
maintenance and operation of the Ryle Telescope, which is supported by
the PPARC.  We also thank the referee for prompt and useful comments.
The Green Bank Interferometer is a facility of the National Science
Foundation and is currently operated by the National Radio Astronomy
Observatory in support of the NASA High Energy Astrophysics programs.
Radio astronomy at the Naval Research Laboratory is supported by the
Office of Naval Research. We acknowledge the use of quick-look results
provided by the ASM/RXTE team.  RPF was supported during the period of
this research initially by ASTRON grant 781-76-017 and subsequently by
EC Marie Curie Fellowship ERBFMBICT 972436.  DJM acknowledges support
for his research by the European Commission under TMR-LSF contract
No. ERBFMGECT950012.


\begin{thebibliography}{}

\bibitem[]{}
Atoyan A.M., Aharonian F.A., 1997, ApJ, 490, L149

\bibitem[]{}
Bailyn C.D., Orosz J.A., McClintock J.E., Remillard R.A., 1995,
Nature, 378, 157

\bibitem[]{}
Bandyopadhyay R., Martini P., Gerard E., Charles P.A., Wagner R.M.,
Shrader C., Shahbaz T., Mirabel I.F., 1998, MNRAS, 295, 623

\bibitem[]{Belloni1}
Belloni T., Mendez M., King A.R., van der Klis M., van Paradijs J.,
1997a, ApJ, 479, L145

\bibitem[]{Belloni2}
Belloni T., Mendez M., King A.R., van der Klis M., van Paradijs J.,
1997b, ApJ, 488, L109


\bibitem[]{CT}
Castro-Tirado A., Brandt S., Lund N., 1992, IAU Circ 5590

\bibitem[]{}
Cawthorne T.V., Wardle J.F.C., Roberts D.H., Gabuzda D.C., 1993,
ApJ, 416, 519

\bibitem[]{}
Downes D., Wilson T.L., Bieging J., Wink J., A\&AS, 1980, 40, 379

\bibitem[]{Eiken}
Eikenberry S.S., Matthews K., Morgan E.H., Remillard R.A., Nelson R.W.,
1998, ApJ, 494, L61

\bibitem[]{}
Feldt M., Stecklum B., Henning Th., Hayward T.L., Lehmann Th., Klein
R., 1998, A\&A, 339, 759

\bibitem[]{FPBN}
Fender R.P., Pooley G.G., Brocksopp C., Newell S.J., 1997, MNRAS, 290, L65

\bibitem[]{}
Fender R.P., Pooley G.G., 1998, MNRAS, 300, 573

\bibitem[]{}
Foster R. S., Waltman E. B., Tavani M., Harmon B. A., Zhang S. N.,
Paciesas W. S., and Ghigo F. D. 1996, ApJ, 467, L81




\bibitem[]{}
Harmon B.A., Deal K.J., Paciesas W.S., Zhang S.N., Robinson C.R.,
Gerard E., Rodr\'\i guez L.F., Mirabel I.F., 1997, ApJ, 477, L85

\bibitem[]{}
Harris D.E., Carilli C.L., Perley R.A. 1994, Nature, 367, 713


\bibitem[]{}
Hjellming R.M., Rupen M.P., 1995, Nature, 375, 464

\bibitem[]{}
Hughes P.A. (editor), 1991, Beams and Jets in Astrophysics, CUP

\bibitem[]{}
Kemball A.J., Diamond P.J., Pauliny-Toth I.I.K., 1996, ApJ, 464, L55



\bibitem[]{}
Levine A.M., Bradt H., Cui W., Jernigan J.G., Morgan E.H.,
Remillard R.A., Shirey R., Smith D., 1996, ApJ, 469, L33

\bibitem[]{}
Longair M.S., 1994, High Energy Astrophysics, Vol 2, 2nd edition, CUP



\bibitem[]{MR}
Mirabel I.F., Rodr\'\i guez L.F., 1994, Nature, 371, 46 [MR94]

\bibitem[]{}
Mirabel I.F. et al., 1994, A\&A, 282, L17

\bibitem[]{}
Mirabel I.F., Rodr\'\i guez L.F., Chaty S., Sauvage M., Gerard E.,
Duc P.-A, Castro-Tirado A., Callanan P., 1996, ApJ, 472, L111 

\bibitem[]{M98}
Mirabel I.F., Dhawan V., Chaty S., Rodr\'\i guez L.F., Mart\'\i {} J.,
Robinson C.R., Swank J., Geballe T.R., 1998, A\&A, 330, L9 



\bibitem[]{}
Patnaik A.R., Browne I.W.A., Wilkinson P.N., Wrobel J.M., 1992, MNRAS 254, 655

\bibitem[]{Pooley1}
Pooley G.G., 1995, I.A.U. Circ 6269


\bibitem[]{PF97}
Pooley G.G., Fender R.P., 1997, MNRAS, 292, 925 [PF97]
 
\bibitem[]{}
Quirrenbach A., Witzel A., Qian S.J., Krichbaum T., Hummel C.A.,
Alberdi A., 1989, A\&A 226, L1

\bibitem[]{}
Rodr\'\i guez L.F., Gerard E., Mirabel I.F., Gomez Y., \& Velazguez A.,
1995, ApJ Supp., 101,173,


\bibitem[]{}
Rodr\'\i guez L.F., Mirabel I.F., 1999, ApJ, in press [RM99]

\bibitem[]{}
Rupen M.P., Hjellming R.M., Mioduszewski A.J., 1998, IAU Circ. 6938

\bibitem[]{Sams}
Sams B., Eckart A., Sunyaev R., 1996, Nature, 382, 47

\bibitem[]{}
Tingay S.J. et al., 1995, Nature, 374, 141


\bibitem[]{} 
 Vermeulen R.C., Schilizzi R.T., Spencer R.E., Romney J.D., Fejes J.,
1993, A\&A, 270, 177

\end{thebibliography}
\end{document}